# Folded MoS$_2$ layers with reduced interlayer coupling

Andres Castellanos-Gomez[*], Herre S. J. van der Zant, and Gary A. Steele

Kavli Institute of Nanoscience, Delft University of Technology, 2628 CJ Delft, The Netherlands.
a.castellanosgomez@tudelft.nl

We study molybdenum disulfide (MoS$_2$) structures generated by folding single- and bilayer MoS$_2$ flakes. We find that this modified layer stacking leads to a decrease in the interlayer coupling and an enhancement of the photoluminescence emission yield. We additionally find that folded single-layer MoS$_2$ structures show a contribution to photoluminescence spectra of both neutral and charged excitons, which is a characteristic feature of single-layer MoS$_2$ that has not been observed in multilayer MoS$_2$. The results presented here open the door to fabrication of multilayered MoS$_2$ samples with high optical absorption while maintaining the advantageous enhanced photoluminescence emission of single-layer MoS$_2$ by controllably twisting the MoS$_2$ layers.

**Introduction**

Single-layer molybdenum disulfide (MoS$_2$) has emerged as a prospective complementary material to graphene because, unlike graphene, it has a large intrinsic direct bandgap [1-3]. This makes monolayer MoS$_2$ an attractive alternative for applications such as: logic circuits [4-6], photodetectors [7-14], light emitters [15-17] and solar cells [18, 19]. Recent studies on the optoelectronic properties of single-layer MoS$_2$ have also shown interesting phenomena including the photogeneration of charged excitons [20] and valley-selective circular dichroism [21-23]. Nevertheless, the case of multilayer MoS$_2$ is different as it is an indirect bandgap semiconductor [24-28], making it less attractive for optoelectronic applications. The direct bandgap nature of single-layer MoS$_2$ is very fragile as the interlayer coupling between just two MoS$_2$ layers is strong enough to turn the bilayer MoS$_2$ into an indirect bandgap semiconductor [24, 25]. This thickness dependent direct-to-indirect bandgap transition has strong consequences in the photoluminescence emission of MoS$_2$: while single-layer MoS$_2$ presents a large photoluminescence yield, the multilayer MoS$_2$ photoluminescence emission is quenched [24, 25].

The aim of this work is to generate artificial MoS$_2$ layered structures with modified interlayer coupling and bandstructure. In this communication we study MoS$_2$ structures where the layers are twisted leading to a decrease in the interlayer coupling and thus an enhancement of the photoluminescence emission yield. Furthermore, we observe that in MoS$_2$ structures formed by two or three twisted MoS$_2$ monolayers, both neutral and charged excitons contribute to the photoluminescence. This feature, typically observed in single layer MoS$_2$, has not been observed in multilayer MoS$_2$ with perfect stacking. Our results demonstrate the potential of twisted MoS$_2$ layers to fabricate optically active materials (similar to single layer MoS$_2$) with high optical absorption (overcoming the main limitation of monolayer MoS$_2$, the low absorption due to its reduced thickness).

**Experimental**

*Sample fabrication*

Twisted MoS$_2$ structures can be fabricated by folding MoS$_2$ layers as they present excellent mechanical properties that allows one to deform the MoS$_2$ layers in extreme ways without breaking them [29-31]. In order to generate folds in the MoS$_2$ layers we transfer MoS$_2$ flakes by mechanical exfoliation onto an elastic substrate (*Gelfilm®* by *Gelpak*) previously stretched by about 100%. Subsequently the strain is suddenly released thereby generating wrinkles in the MoS$_2$ layers through buckling-induced delamination [32, 33]. Interestingly, for single- and bilayer MoS$_2$ these wrinkles are not stable due to their reduced bending rigidity and they tend to collapse forming bifolds (see the sketch in Figure 1a and the optical image in Figure 1c). We also found that the edges of single-layer flakes tend to fold after the sample fabrication process (see Figure 1a and Figure 1d). We consider this approach to induce cleaner folded structures than other techniques previously used to generate folded graphene such as water flushing the fabricated samples to induce the folds [34, 35]. In that case the water flushing may lead to adsorbated layers between the folded layers, hampering the interpretation of the data. Moreover these techniques, that rely on partially folding the edge of the flakes, can only fabricate simply folded structures. Our technique also offers the possibility of fabricating bifolded structures by collapse of the wrinkles induced during the fabrication process.

In the folded regions, the MoS$_2$ layers are not stacked following the normal AB stacking, *i.e.*, the MoS$_2$ layers are twisted (see Figure 1b and the Electronic Supporting Material). Using this fabrication process, we have fabricated folded single layers (hereafter denoted by 1L+1L), bifolded single-layers (1L+1L+1L) and bifolded bilayers (2L+2L+2L). For our experiments, we use selected folds and bifolds wider than 2 μm to ensure that the laser spot (400 nm in diameter) only probes the twisted regions, and also avoid laser spot positions close to the edges of the folds where the mechanical strain can alter the optoelectronic properties of MoS$_2$ [32, 36-42].

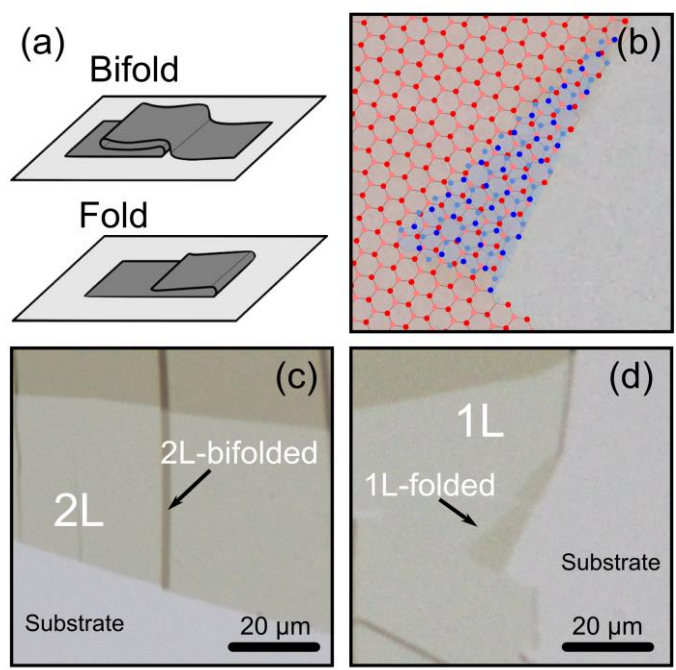

**Figure 1** (a) Sketches of the MoS$_2$ structures fabricated by folding twice an MoS$_2$ layer (bifold) and by folding one edge (fold). (b) Cartoon illustrating how the folding process typically results in a MoS$_2$ structure composed of twisted layers that are not perfectly AB stacked. (c), (d) Optical microscopy images of a bifolded MoS$_2$ bilayer and a folded single-layer MoS$_2$ respectively.

*Optical microscopy*

MoS$_2$ sheets have been identified under an optical microscope (*Olympus BX 51* equipped with a *Canon EOS 600D* camera) [43]. As the employed elastic substrates are transparent, one can employ transmission mode optical microscopy to determine the number of MoS$_2$ layers. In a previous work we found that the transmittance ($T$) of MoS$_2$ flakes up to ten layers thick depends on the number of layers ($N$) as $T = 100 - 5.5N$ (see the Electronic Supplementary Material of Ref. [32]). We found that the transmittance measured for folded (1L+1L) and bifolded (1L+1L+1L and 2L+2L+2L) samples is in good agreement with the transmittance measured for pristine (perfectly AB stacked) 2L, 3L and 6L respectively (see Figure S4 in the ESM).

*Raman spectroscopy and photoluminescence*

A *Renishaw in via* system was employed to perform simultaneous micro-Raman spectroscopy and photoluminescence measurements on the fabricated twisted MoS$_2$ structures. The system was used in a backscattering configuration excited with visible laser light ($\lambda$ = 514 nm). The Raman and photoluminescence spectra were simultaneously collected through a 100× objective (NA = 0.95) and recorded with 1800 lines/mm grating providing the spectral resolution of ~ 1 cm$^{-1}$. To avoid laser-induced modification or ablation of the samples, all spectra were recorded at P ~ 500 μW which is found to be low enough power to prevent laser-induced damage [44-46]. Note that the use of *Gelfilm®* substrate provides an enhanced photoluminescence signal with respect to conventionally used SiO$_2$/Si substrates [47] which is advantageous for photoluminescence studies. Moreover, the photoluminescence spectra of samples fabricated onto *Gelfilm®* presents reduced doping level in comparison to samples prepared on SiO$_2$. This reduced doping level yields photoluminescence spectra very similar to that of MoS$_2$ samples on boron nitride or freely suspended (See Figure S2 of the ESM).

### Results and discussion

Raman spectroscopy has proven to be a powerful tool to characterize atomically thin materials [48] and it has been previously used to characterize atomically thin MoS$_2$ layers [49]. In the Raman spectrum of MoS$_2$ there are two prominent peaks corresponding to the $E^1_{2g}$ and the $A_{1g}$ vibrational modes. In the $E^1_{2g}$ mode, the Mo and S atoms oscillate in anti-phase and parallel to the surface plane (see scheme in the inset in Figure 2a). In the $A_{1g}$ mode, the S atoms oscillate in anti-phase perpendicularly to the surface plane while the Mo atoms remain static. It has been shown that the frequency difference between these two Raman modes increases monotonically with the number of layers [49, 50]. For single-layer MoS$_2$, which is not subjected to any interlayer interaction, the frequency difference between the $E^1_{2g}$ and the $A_{1g}$ peaks is around 19 cm$^{-1}$. For perfectly AB stacked bilayer MoS$_2$ this value is about 21 cm$^{-1}$. For bulk MoS$_2$, the layers feel the interaction of many other MoS$_2$ layers yielding a frequency difference of 25.5 cm$^{-1}$. Therefore, the frequency difference between the $E^1_{2g}$ and the $A_{1g}$ peaks can also be used to estimate the strength of the interlayer coupling.

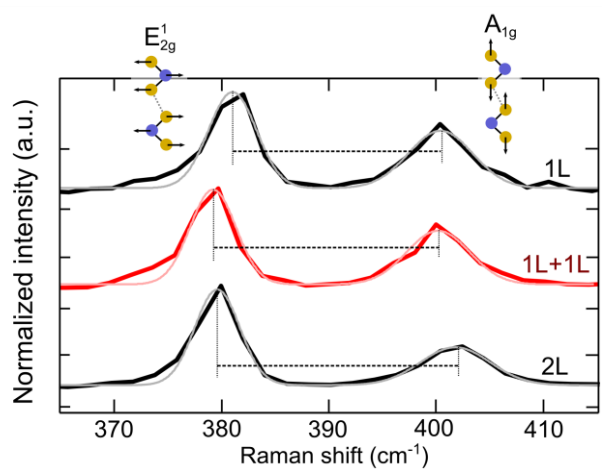

**F**igure 2 Raman spectra measured for folded single-layer (1L+1L, in red) MoS$_2$ layers. The Raman spectra measured for pristine 1L and 2Lhave been included for comparison. The frequency difference between the $E^1_{2g}$ and $A_{1g}$ Raman modes is indicated with dashed horizontal lines. The thin light lines are Lorentzian fits to the experimental data.

In Figure 2, the Raman spectrum obtained for folded monolayer MoS$_2$ (1L+1L) is compared with results acquired for pristine single- and bilayer MoS$_2$ because they can be considered as two limiting cases: a system without interlayer coupling and one with the interlayer coupling due to the perfect AB stacking of the two MoS$_2$ layers. The folded single-layer MoS$_2$ shows a frequency difference between the E$^1_{2g}$ and the A$_{1g}$ that lies in between the one measured for 1L and 2L MoS$_2$. This indicates that the 1L+1L layer has a reduced interlayer coupling with respect to the perfectly stacked 2L MoS$_2$. We also systematically observed blueshift of the E$^1_{2g}$ that may be due to a change in the strain of the layer[51, 52]. Similar results have been found for bifolded single-layer (1L+1L+1L) and bifolded bilayer (2L+2L+2L) MoS$_2$ (see Electronic Supplementary Material).

For all the studied cases (1L+1L, 1L+1L+1L and 2L+2L+2L), the frequency difference between the Raman modes lies between the limiting cases. This indicates that the twisted MoS$_2$ layers exhibit a reduced interlayer coupling in comparison to the case of perfectly stacked layers. The reduction of the interlayer coupling, however, is not enough to consider them as completely independent layers. Figure 3 summarizes the measured frequency difference between the E$^1_{2g}$ and the A$_{1g}$ Raman modes for pristine layers (perfect stacking) and for 1L+1L, 1L+1L+1L and 2L+2L+2L MoS$_2$ layers. The frequency difference is systematically lower for the folded layers than for their perfectly stacked counterparts. Nonetheless, we have found dispersion on the values measured for different folded and bifolded layers which we attribute to difference in the twisting angle which may lead to different interlayer coupling.

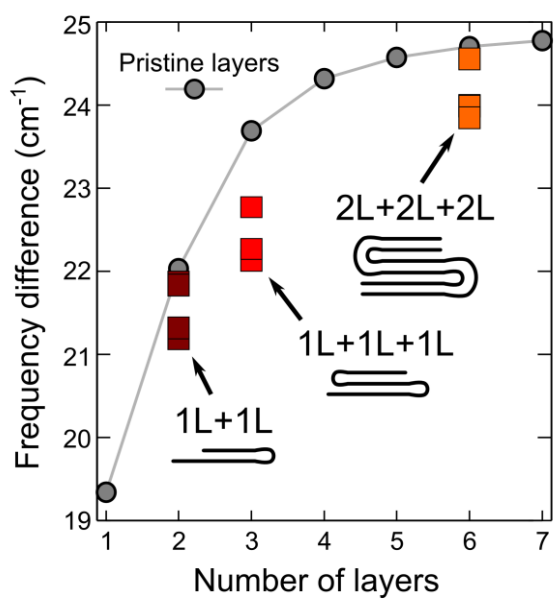

**Figure 3** Frequency difference between the E$^1_{2g}$ and A$_{1g}$ Raman modes as a function of the number of layers measured for: pristine layers (gray circles), folded and bifolded single-layer (1L+1L and 1L+1L+1L, dark and light red squares respectively) and bifolded bilayer MoS$_2$ (2L+2L+2L, orange squares).

Photoluminescence measurements have been used to study the bandstructure of atomically thin MoS$_2$ [24-26, 53]. The photoluminescence spectrum of MoS$_2$ presents prominent peaks around 630 nm and 670 nm which correspond to the B and A excitons. For single-layer MoS$_2$, the A exciton peak is composed of two peaks corresponding to the contribution of both charged (labeled as A$^-$, occurring at ~ 655 nm) and neutral excitons (labeled as A, occurring at ~ 670 nm) to the photoluminescence emission. On the other hand, multilayered MoS$_2$ does not show a contribution due to charged excitons yielding a single A exciton peak. Furthermore, multilayer MoS$_2$ presents an extra peak at lower energy (between 790 nm and 1000 nm, depending on the number of layers) corresponding to the indirect bandgap transition (labeled as I).

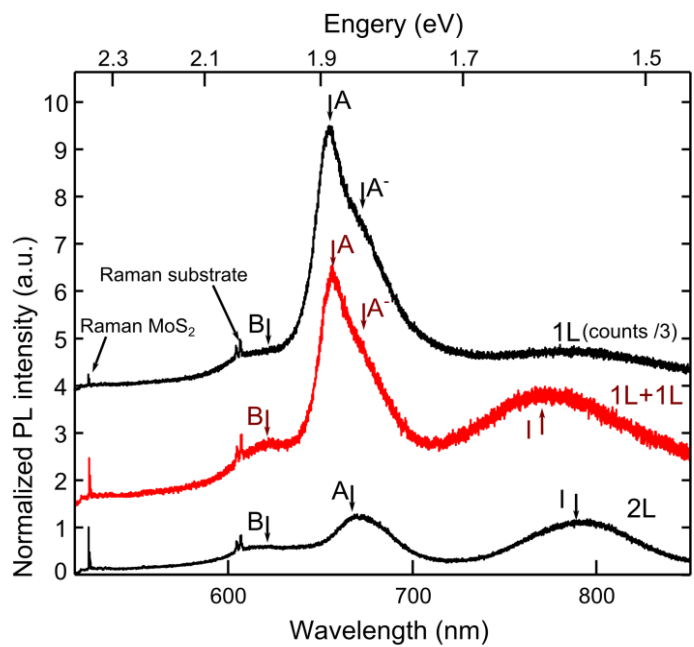

**Figure 4** Photoluminescence spectra measured for single-layer (1L), folded single-layer (1L+1L) and bilayer (2L) MoS$_2$. The peaks corresponding to the: Raman emission of the MoS$_2$ and substrate, the B exciton, the A and A$^-$ excitons and the I exciton have been highlighted.

Figure 4 shows the spectrum measured for folded single-layer MoS$_2$ compared to the limiting cases (1L and 2L MoS$_2$). Similar measurements for bifolded single- and bilayer MoS$_2$ can be found in the Electronic Supplementary Material. The photoluminescence spectra have been normalized to the E$^1_{2g}$ Raman peak intensity to account for the fact that thicker MoS$_2$ layers absorb more excitation light as the photoluminescence yield is expected to be proportional to the absorbed excitation (See Figure S3 of the ESM). The folded MoS$_2$ monolayer shows an enhanced photoluminescence emission of the A exciton in comparison to pristine 2L MoS$_2$ (1.7 to 4.1 times larger) [54]. Moreover, the A exciton peak is composed of two Lorentzian peaks, indicating that both neutral and charged excitons contribute to the photoluminescence emission, similarly to the case of pristine 1L MoS$_2$ [20] (See the Electronic Supporting Information for more detailed analysis of the photoluminescence spectra). These two observations (enhanced photoluminescence and a contribution of charged excitons to the photoluminescence yield) indicate that a twisted MoS$_2$ bilayer presents a reduced interlayer coupling with respect to perfectly AB stacked bilayer MoS$_2$, a conclusion consistent with the one obtained from the Raman studies. Moreover, the folded single-layer MoS$_2$ exhibits an indirect bandgap transition with higher energy than that of pristine bilayer MoS$_2$, as indicated by the blueshift (20 nm) of the I excitonic peak.

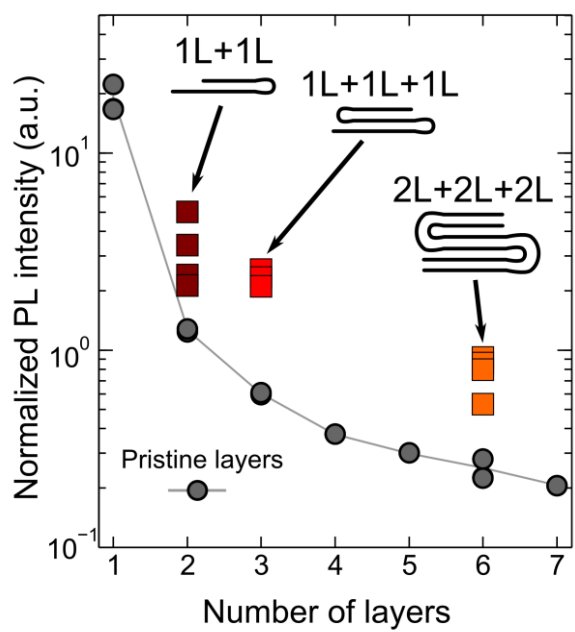

**Figure 5** Normalized photoluminescence intensity of the A exciton peak (at 670 nm) as a function of the number of layers, measured for pristine layers (gray circles) and folded single-layer (dark red squares), bifolded single-layer (red squares) and bifolded bilayer (orange squares).

We have found that for all studied samples (four 1L+1L, three 1L+1L+1L and five 2L+2L+2L samples), the photoluminescence intensity always show an enhancement of up to a factor 5 in comparison to their perfectly stacked counterparts (see Figure 5). This result further proof that folded and bifolded MoS$_2$ layers exhibit a reduced interlayer coupling, in agreement with the conclusions obtained from the Raman spectroscopy and the analysis of the excitonic peaks of the photoluminescence spectra of folded single-layer MoS$_2$.

### Conclusions

In summary, MoS$_2$ structures with twisted layers have been fabricated by folding pristine layers via an all dry process involving pre-stressed elastic substrates. Raman spectroscopy measurements indicate that the interlayer coupling strength is reduced due to the twisting. The photoluminescence spectra of twisted 1L (both folded and bifolded 1L) shows the contribution of neutral and charged excitons: the presence of charged excitons has only been observed in single-layer MoS$_2$. Moreover, the photoluminescence spectra of twisted layers presents a peak corresponding to an indirect bandgap transition with smaller energy than that expected for perfectly stacked layers. In addition, photoluminescence emission yield is larger than that of perfectly stacked layers (up to a factor of 5). These observations all point at a modification of the interlayer coupling, and thus the bandstructure, by altering the natural stacking in MoS$_2$.

More specifically, the interlayer coupling is reduced with respect to layers with optimal coupling and in many aspects twisted layers behave more like single-layer $MoS_2$. The results shown here open the door to design artificial 3D $MoS_2$ materials whose optoelectronic properties are engineered by the twisting angle between the $MoS_2$ layers. Such 3D structures may have applications in photodetection and photovoltaics as one could fabricate multilayered $MoS_2$ devices with high optical absorption (potentially much higher than that of monolayer $MoS_2$) while maintaining some of the advantageous optoelectronic properties of single-layer $MoS_2$.

## Acknowledgements

The authors like to acknowledge fruitful discussions with J. Fernández-Rossier (INL, Portugal), A.C. Ferrari, R. S. Sundaram (Cambridge University, UK), R. Roldán, P. San-Jose (ICMM-CSIC, Spain) and E. Prada (Universidad Autonoma de Madrid). This work was supported by the European Union (FP7) through the program RODIN. A.C.-G. acknowledges financial support through the FP7-Marie Curie Project PIEF-GA-2011-300802 ('STRENGTHNANO').

**Electronic Supplementary Material**

# Folded MoS$_2$ layers with reduced interlayer coupling

Andres Castellanos-Gomez[*], Herre S. J. van der Zant, and Gary A. Steele

Kavli Institute of Nanoscience, Delft University of Technology, 2628 CJ Delft, The Netherlands.
a.castellanosgomez@tudelft.nl

**Table of contents**

---

Address correspondence to: Andres Castellanos-Gomez, a.castellanos-gomez@tudelft.nl

This is the post-peer reviewed version of the following article:
A.Castellanos-Gomez *et al*. "Folded MoS₂ layers with reduced interlayer coupling".
Nano Research, 2014, 7(4): 572–578
DOI 10.1007/s12274-014-0425-z
Published in its final form at:
http://link.springer.com/article/10.1007%2Fs12274-014-0425-z**Transfer of the folded and bi-folded MoS₂ samples to other substrates**

After fabrication of the folded and bi-folded MoS₂ structures onto a Gelfim substrate, the samples can be transferred to a different substrate. The surface of the Gelfilm substrate is pressed against the new substrate and peeled off very slowly, transferring some of the flakes. We found that the slower the peeling off the higher transfer yield. Figure S1 shows an example of a large area 2 to 8 layers thick flake with many bifolds which has been partially transferred onto a SiO₂/Si substrate.

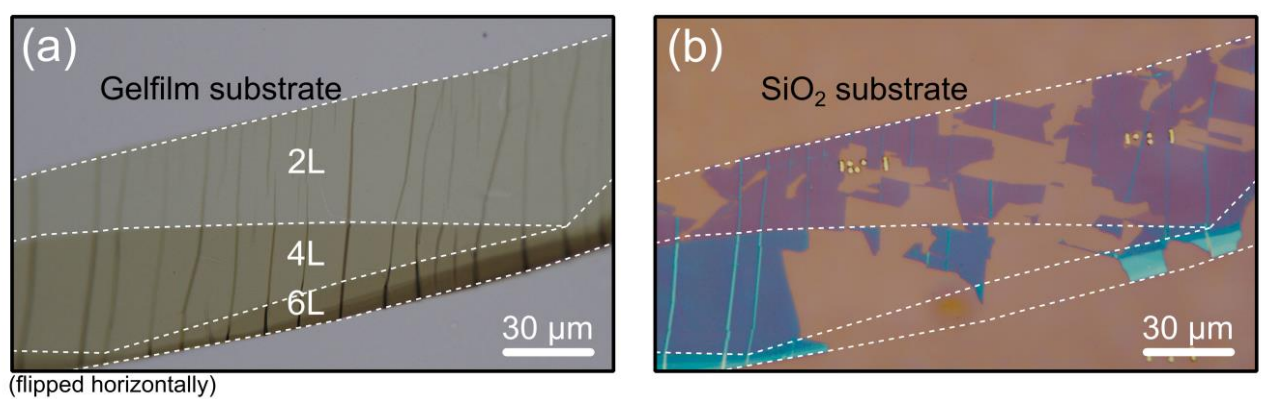

**Figure S1** Optical images of a bifolded MoS₂ sample, as fabricated on a Gelfilm substrate (a) and after transfer part of the flake onto a SiO₂/Si substrate.

**Doping of MoS₂ samples on Gelfilm and SiO₂ substrates**

The photoluminescence experiments shown in the main text have been carried out on samples fabricated onto Gelfilm because of their lower doping level. Indeed, we recently found that the photoluminescence spectra of samples fabricated onto Gelfilm presents reduced doping level in comparison to samples prepared on SiO₂. This reduced doping level yields photoluminescence spectra very similar to that of freely suspended MoS₂. Figure S2 shows a comparison of the photoluminescence spectra measured in single-layer MoS₂ flakes deposited onto Gelfilm (panel a) and SiO₂ (panel b). The contribution of the different excitons to the photoluminescence can be determined by performing a multi-Gaussian fit to the data. The different excitons have been highlighted on the plot, being very clear the low intensity of the neutral A exciton in the sample prepared on SiO₂, which indicates a high doping level on the flake. The samples prepared on Gelfilm, on the other hand, present an almost equal ratio of charged and neutral excitons, very similar to what is observed on freely suspende MoS₂.

**Normalization of the photoluminescence spectra**

In order to compare the intrinsic photoluminescence yield of MoS₂ samples with different thicknesses one has to account for the fact that thicker samples will absorb more excitation light and their photoluminescence intensity will be larger. In order to discount for this effect (to determine the intrinsic photoluminescence yield) one can normalize the photoluminescence spectra with a normalization value that depends on the absoption. In the literature the intensity of some of the Raman peaks is typically employed to account this effect. Figure S3a shows several Raman spectra measured for MoS$_2$ samples with different thicknesses under similar conditions of incident power, acquisition time and focusing. The intensity of the Raman peaks depends linearly on the number of layers within experimental uncertainty (Figure S3b) and therefore it can be used to normalize the photoluminescence spectra.

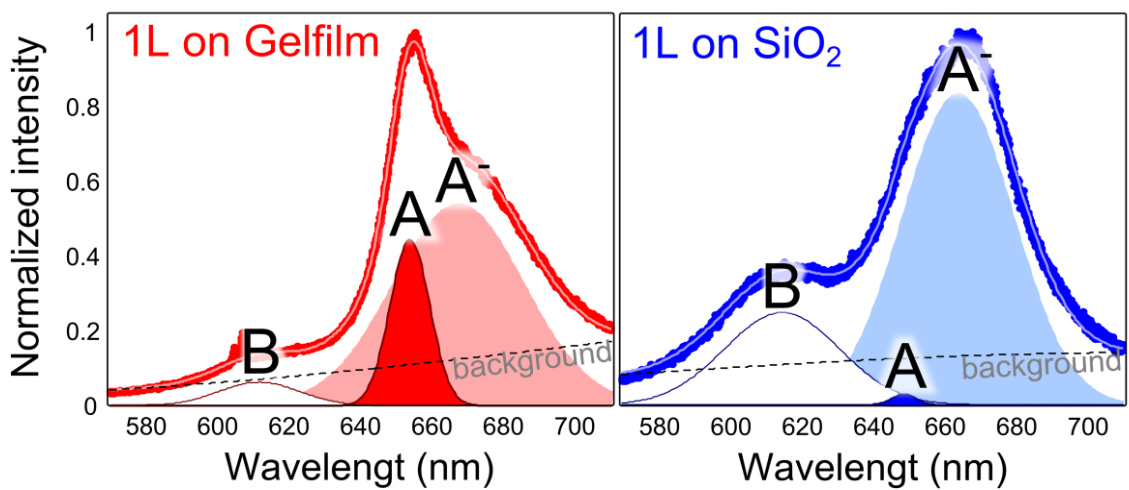

**Figure S2** Photoluminescence spectra of single-layer MoS$_2$ samples fabricated on Gelfilm and SiO$_2$/Si substrates. A multiple-Gaussian fit has been used to determine the intensities of the B, A, and A$^-$(trion) excitons. The neutral A exciton intensity is very small for samples fabricated on SiO$_2$, due to high doping level. On the other hand, samples fabricated onto Gelfilm present a PL spectrum very similar to freely suspended MoS$_2$ with high intensity of the A neutral exciton.

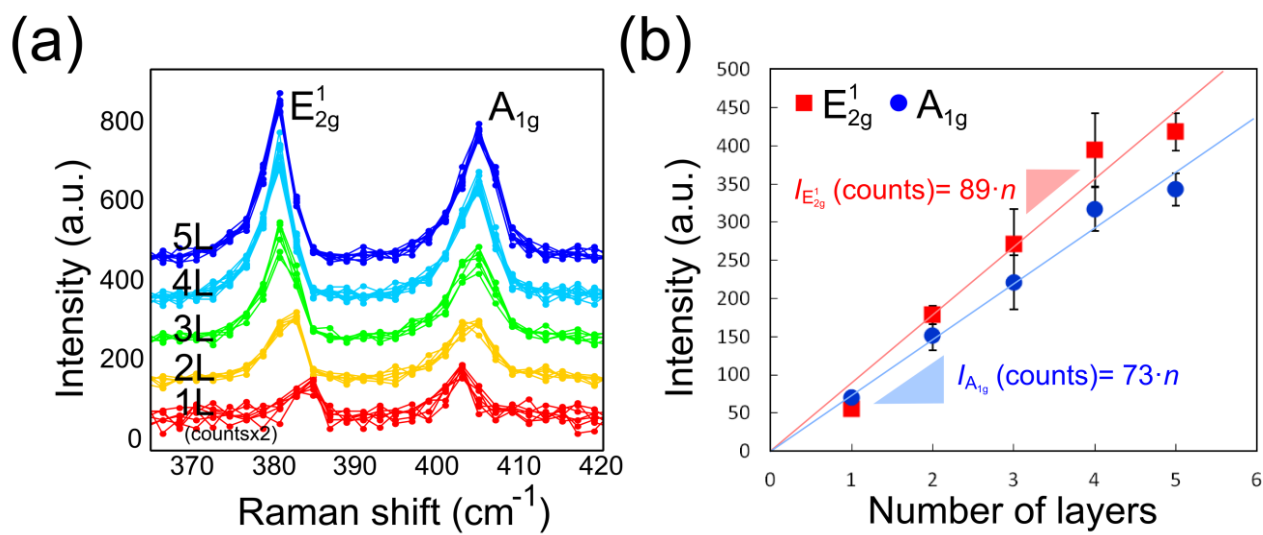

**Figure S3** (a) Raman spectra measured for MoS₂ samples with different number of layers fabricated onto Gelfilm substrates, measured under same intensity, focus and acquisition time conditions. (b) Dependence of the intensity of the Raman peaks with the number of layers, showing that the intensity is proportional to the number of layers.

**Sample characterization**

Apart from the Raman spectroscopy and photoluminescence characterization, the fabricated samples were studied by quantitative optical microscopy and atomic force microscopy. As the samples were fabricated onto Gelfilm substrates, which are transparent, they can be studied in transmission mode optical microscopy. Figure S4a shows a false color map of the transmittance of a MoS₂ sample with a folded monolayer area. In order to obtain the transmittance, an optical image is acquired using a digital camera attached to the trinocular of the microscope. The values of the red, green and blue channels are summed to make a matrix of intensities per pixel. Especial attention is taken to avoid using images with some channel saturated or underexposed. Then the transmittance is obtained by dividing the matrix by the average intensity on the bare substrate. The resulting matrix can be plotted in a false color image as Figure S4a. Figure S4b presents the transmittance values measured for both pristine and folded/bifolded samples with different thicknesses. The transmittance of the folded/bifolded samples is in good agreement with the values obtained for pristine samples.

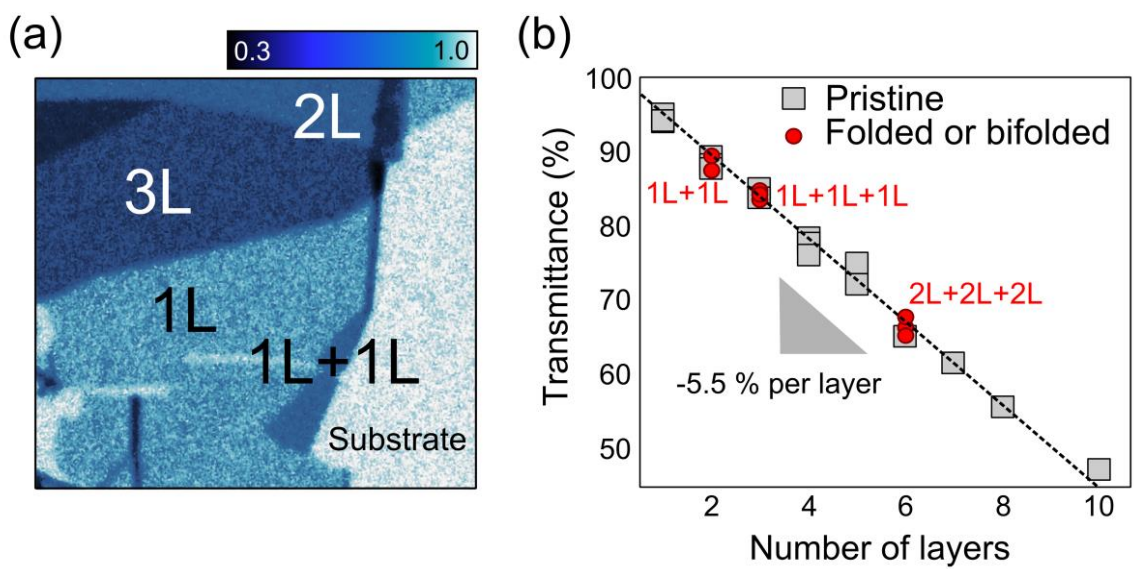

**Figure S4** (a) Optical transmittance map of a folded MoS$_2$ sample. (b) Dependence of the transmittance with the number of layers, included datapoints measured on folded and bifolded layers.

Atomic force microscopy has been used to study the topography of the fabricated samples. However, this characterization is very challenging for the samples fabricated on Gelfilm substrates as the substrate easily deforms during the scanning leading to setpoint-dependent images. Moreover, the large tip-Gelfilm interaction makes the feedback loop more unstable on the regions uncovered by the MoS$_2$. Therefore, studying the details of the folded structures results non-trivial. To overcome this issue one can transfer the folded/bifolded structures onto a flat SiO$_2$/Si surface before characterizing with the AFM. Figure S5 shows two AFM images of bifolded bilayer MoS$_2$ structures as fabricated on Gelfilm and after transfer to a SiO$_2$ surface. From the image in Figure S5b one can see that the bifolded structure is composed by twisted bilayers of MoS$_2$ stacked on top of each other.

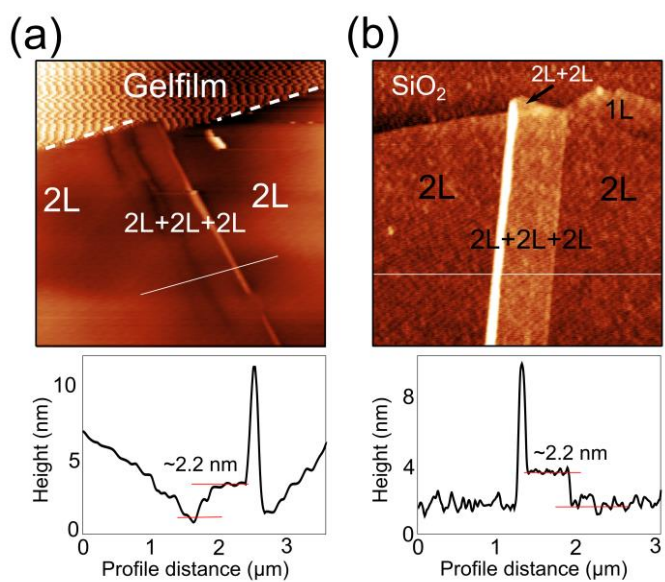

**Figure S5** Atomic force microscopy images of bifolded bilayer MoS$_2$ on gelfilm (a) and SiO$_2$/Si (b) substrates. Topographic line profiles, measured along the solid white lines, are included below the images.

## Raman and PL spectra measured for different folded MoS₂ structures

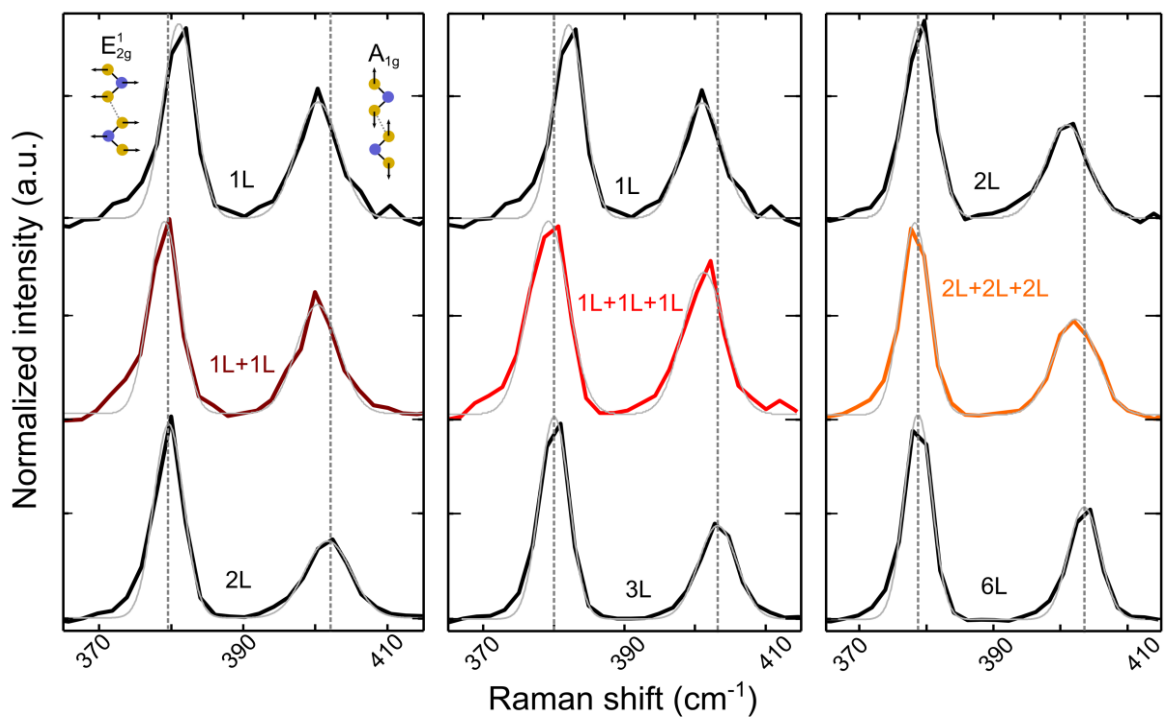

**gure S6** Raman spectra measured for folded (1L+1L) and bifolded (1L+1L+1L) monolayer MoS₂ and bifolded bilayer MoS₂ (2L+2L+2L). Measurements on pristine 1L, 2L, 3L and 6L MoS₂ have been also included to facilitate the comparison as they can be considered as limiting cases: negligible and optimal interlayer coupling. The thin light lines are Lorentzian fits to the experimental data.

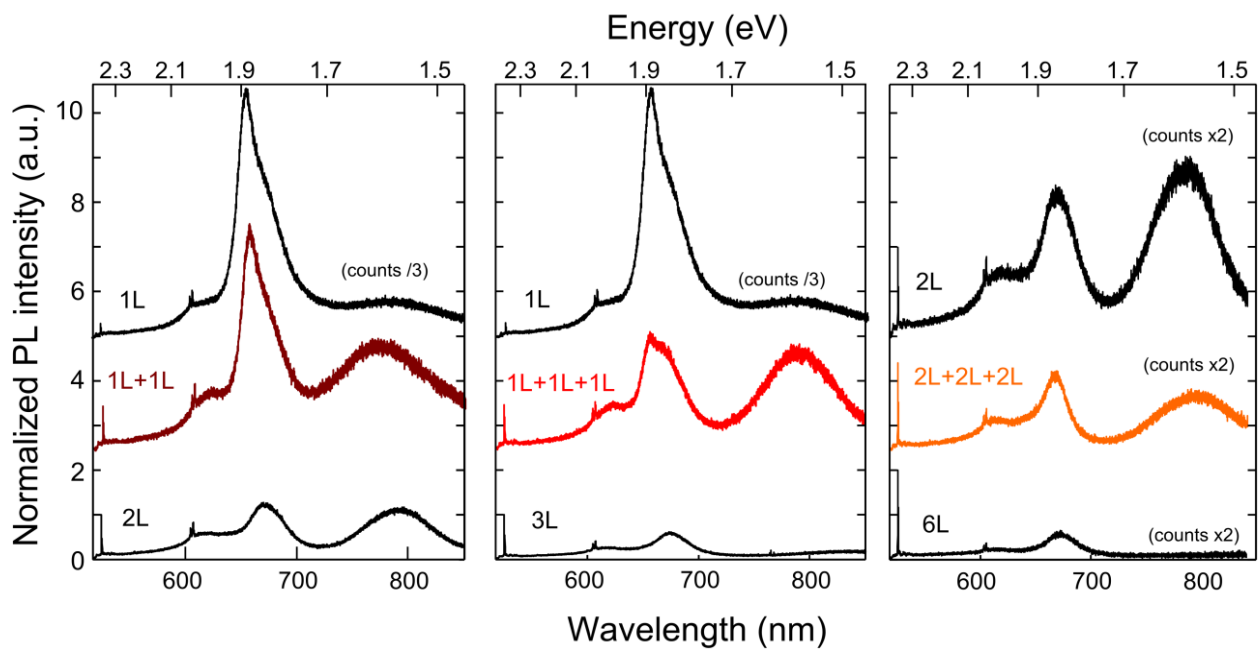

**Figure S7** Photoluminescence spectra measured for folded (1L+1L) and bifolded (1L+1L+1L) monolayer MoS$_2$ and bifolded bilayer MoS$_2$ (2L+2L+2L). Measurements on pristine 1L, 2L, 3L and 6L MoS$_2$ have been also included to facilitate the comparison as they can be considered as limiting cases: negligible and optimal interlayer coupling.

**Analysis of the photoluminescence spectra measured for different folded MoS$_2$ structures**

Similarly to the results presented in the main text for the A exciton, the B exciton photoluminescence emission is also enhanced for folded and bifolded samples in comparison with their pristine counterparts. Figure S8 shows a plot with the comparison of the B exciton intensity (normalized to the E$^1_{2g}$ Raman peak) as a function of the number of layers. The folded and bifolded structures present B exciton intensities larger than the pristine layers (up to a factor of 5), indicating a reduced interlayer coupling due to the folded structure.

Table S1 summarizes the normalized intensities of the different peaks composing the A exciton (the neutral A exciton and the charged A$^-$ exciton) determined by performing multi-Gaussian fits as those shown in Figure S2. For pristine monolayer the neutral A exciton occurs at ~655 nm and the charged occurs at ~670 nm. For multilayers, however, the emission is composed exclusively by one peak around ~670 nm associated to the emission of neutral excitions in multilayer MoS$_2$ (in agreement with the lack of observation of trions in multilayer MoS$_2$). Folded and bifolded single-layer MoS$_2$ behave very differently from their pristine counterparts as their PL emission shows again two peaks exactly at the wavelengths associated to the emission of neutral and charged excitons in single-layer MoS$_2$. This indicates than these structures formed by folded and bifolded single-layers behave similarly to pristine single-layer MoS$_2$. For bifolded bilayer, the PL spectra only presents one peak around ~670 nm.

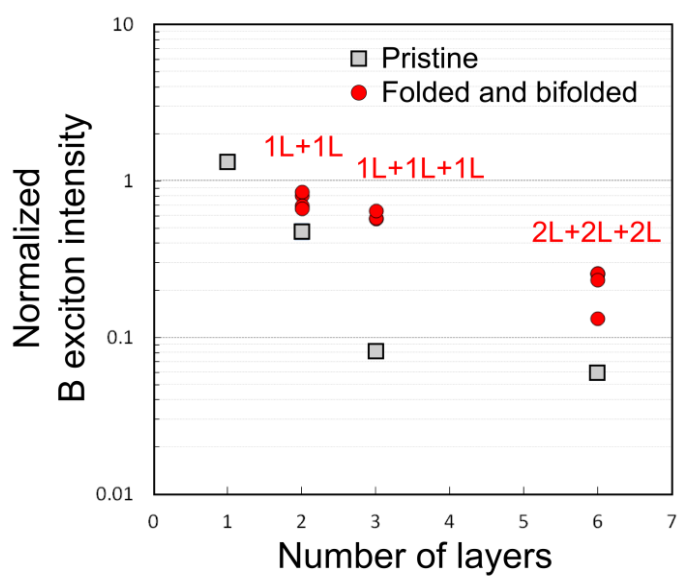

**Figure S8** Normalized intensity of the B exciton peak as a function of the number of layers. Datapoints measured for folded and bifolded structures have been also included. Similarly to what is showed for A exciton (Figure 4 of the main text) the B exciton emission yield is enhanced for folded and bifolded structures.

**Table S1** Summary of the normalized intensity of the peaks appearing around 655 nm and 670 nm for pristine and folded/bifolded structures with different thicknesses.

| Pristine structures | | | Folded or bifolded structures | | |
| --- | --- | --- | --- | --- | --- |
| # of layers | $I_{Peak}$ at ~655nm | $I_{Peak}$ at ~670nm | # of layers | $I_{Peak}$ at ~655nm | $I_{Peak}$ at ~670nm |
| 1L | 6.81 | 9.83 | 1L + 1L | 2.00 | 2.82 |
| 2L | 0.0 | 1.24 | 1L + 1L | 0.82 | 2.40 |
| 3L | 0.0 | 0.60 | 1L + 1L | 0.95 | 1.86 |
| 4L | 0.0 | 0.38 | 1L + 1L | 0.23 | 1.73 |
| 5L | 0.0 | 0.30 | 1L + 1L +1L | 0.57 | 1.64 |
| 6L | 0.0 | 0.28 | 1L + 1L +1L | 0.65 | 1.83 |
| 6L | 0.0 | 0.22 | 1L + 1L +1L | 0.44 | 1.90 |
| 7L | 0.0 | 0.21 | 2L + 2L + 2L | 0.0 | 0.66 |
| | | | 2L + 2L + 2L | 0.0 | 0.69 |
| | | | 2L + 2L + 2L | 0.0 | 0.80 |
| | | | 2L + 2L + 2L | 0.0 | 0.75 |
| | | | 2L + 2L + 2L | 0.0 | 0.45 |